\documentclass[12pt]{article}
\usepackage{graphicx}
\begin{document}
{\large\bf Spintronic devices on the base of magnetic
nanostructures}

\bigskip
{\large L.V. Lutsev, A.I. Stognij *, N.N. Novitskii *, and A.S.
Shulenkov **}

\bigskip
{\it A.F. Ioffe Physical-Technical Institute, Russian Academy of
Sciences, 194021, St Petersburg, Russia\\ ${ }$\\${ }^*$ Scientific
and Practical Materials Research Center, National Academy of
Sciences of Belarus, 220072 Minsk, Belarus\\${ }$\\${ }^{**}$ Minsk
Research Institute of Radiomaterials, 220074, Minsk, Belarus\\${ }$\\
E-mail: l\_lutsev@mail.ru}

\begin{abstract}
Two types of spintronic devices on the base of magnetic
nanostructures containing silicon dioxide films with cobalt
nanoparticles SiO${ }_2$(Co) on GaAs substrate -- magnetic sensors
and field-effect transistor governed by applied magnetic field --
are studied. Magnetic sensors are based on the injection
magnetoresistance effect. This effect manifests itself in avalanche
suppression by the magnetic field in GaAs near the SiO${
}_2$(Co)/GaAs interface. Field-effect transistor contains the SiO${
}_2$(Co) film under gate. It is found that the magnetic field action
leads to great changes in electron mobility in the channel due to
the interaction between spins of Co nanoparticles and electron
spins.
\end{abstract}

\section{Introduction}
Manipulation of carrier spins in ferromagnetic / semiconductor
heterostructures offers enhanced functionality of spin-electronic
devices such as spin transistors, sensors, and magnetic memory
cells~\cite{Wolf,Schmidt}. This manipulation can be realized on the
base of magnetic nanostructures by use of magnetoresistance effects
and due to interactions between magnetic nanostructures and electron
spins in field-effect transistors. Magnetoresistance effects are
attracting much attention in a view of their various applications.
An extremely large magnetoresistance ($10^5$ \%) has been observed
at room temperature in GaAs / granular film heterostructures in the
avalanche state, where a granular film contains ferromagnetic metal
nanoparticles or ferromagnetic islands on the semiconductor / film
interface. The value of this effect is two-three orders higher than
the maximum values of the giant magnetoresistance in the metal
magnetic multilayers and the tunnelling magnetoresistance in the
magnetic tunnel junction structures. Magnetoresistance effects of
high values have been found in GaAs / granular film heterostructures
with granular films containing ferromagnetic metal MnSb nano-islands
\cite{Akin00,Akin02} and ferromagnetic MnAs clusters~\cite{Yok06}.
The high values of the magnetoresistance based on the avalanche
breakdown has been observed on the SiO${ }_2$(Co)/GaAs
heterostructures, where the SiO${ }_2$(Co) is the granular SiO${
}_2$ film containing Co
nanoparticles~\cite{Lut05,Lut06,Lut09,Lut11}. This effect has been
called the injection magnetoresistance (IMR). It appears, when
electrons are injected from the granular film into the GaAs.

Field-effect HEMT devices with spin polarized electron channels are
perspective due to the possibility to change electron spins in
channels by a magnetic field action. In this paper, we study devices
on the base of magnetic nanostructures with silicon dioxide films
with cobalt nanoparticles SiO${ }_2$(Co) on GaAs substrate --
magnetic sensors and field-effect transistor governed by applied
magnetic field. Magnetic sensors are based on the IMR effect. The
field-effect transistor contains the SiO${ }_2$(Co) film under gate.
The magnetic field action leads to great changes in the electron
mobility in the channel due to the interaction between spins of Co
nanoparticles and electron spins.

\section{Giant injection magnetoresistance and magnetic sensors}

\begin{figure*}
\begin{center}
\includegraphics*[scale=.55]{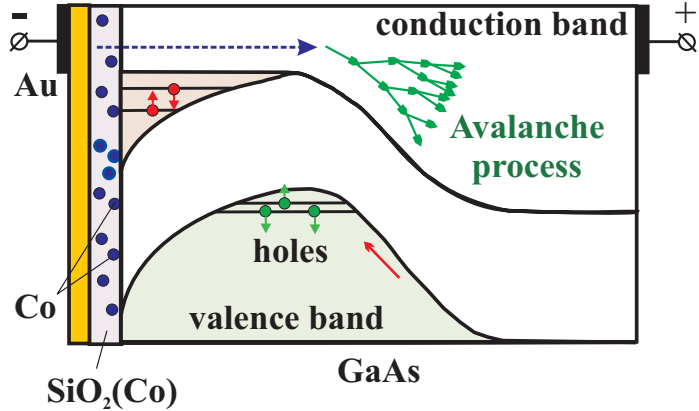}
\end{center}
\caption{Schematic energy band diagram of the magnetoresistive
sensor on the base of the heterostructure with a quantum well near
the interface and a hole trap in the avalanche regime. }
\label{Fig1}
\end{figure*}

High values of the IMR-effect in the SiO${ }_2$(Co) heterostructures
are explained by the magnetic-field-controlled process of impact
ionization in the vicinity of the spin-dependent potential barrier
formed in the semiconductor (Fig. \ref{Fig1}). The spin-dependent
potential barrier is formed near the interface by the exchange
interaction between the electrons in localized states in the
electron accumulation layer in the semiconductor and $d$-electrons
of Co~\cite{Lut09,Lut06a}. The avalanche process is induced by
electrons, which (1) surmount over the spin-dependent potential
barrier formed by the exchange-splitted localized states and (2)
tunnel from the localized states. The impact ionization induced by
the injected electrons produces holes, which move, are accumulated
in the region of the potential barrier and lower the barrier
height~\cite{Lut09}. Owing to the formed hole feedback, small
variations in the barrier height lead to great changes in the
current and in the avalanche process. The applied magnetic field
increases the barrier height, reduces the transparency of the
barrier, and suppresses the onset of the impact ionization.

Magnetic sensors were performed on the samples SiO${ }_2$(Co)/GaAs
with the $n$-GaAs substrates. Carrier concentrations in the $n$-GaAs
are equal to $10^{15}$ cm${ }^{-3}$. The SiO${ }_2$(Co) films were
deposited by the ion-beam co-sputtering of the composite
cobalt-quartz target onto the GaAs substrates heated to
200$^{\circ}$C. The concentration of Co nanoparticles in the silicon
dioxide was varied by changing the ratio of target areas of cobalt
and quartz areas. The film composition was determined by the nuclear
physical methods of element analysis using a deuteron beam of the
electrostatic accelerator (Rutherford backscattering spectrometry
and nuclear reaction with oxygen). For the samples used in magnetic
sensors, the relative Co content is in the range 45 - 71~at.\% and
the film thickness is 40~nm. The average size of Co particles,
determined from the low-angle X-ray scattering measurements,
increases with Co content: from 2.9~nm at 45~at.\% to 3.9~nm at
71~at.\%. As the Co content increases, the resistivity of the SiO${
}_2$(Co) films decreases from 0.3~$\Omega\cdot$cm (45~at.\%) through
3.0$\cdot 10^{-3}$ $\Omega\cdot$cm (60~at.\%) to 1.4$\cdot 10^{-3}$
$\Omega\cdot$cm (71~at.\%). The protective Au layer of the thickness
3 - 5 nm have been sputtered on the SiO${ }_2$(Co) films.

\begin{figure*}
\begin{center}
\includegraphics*[scale=.5]{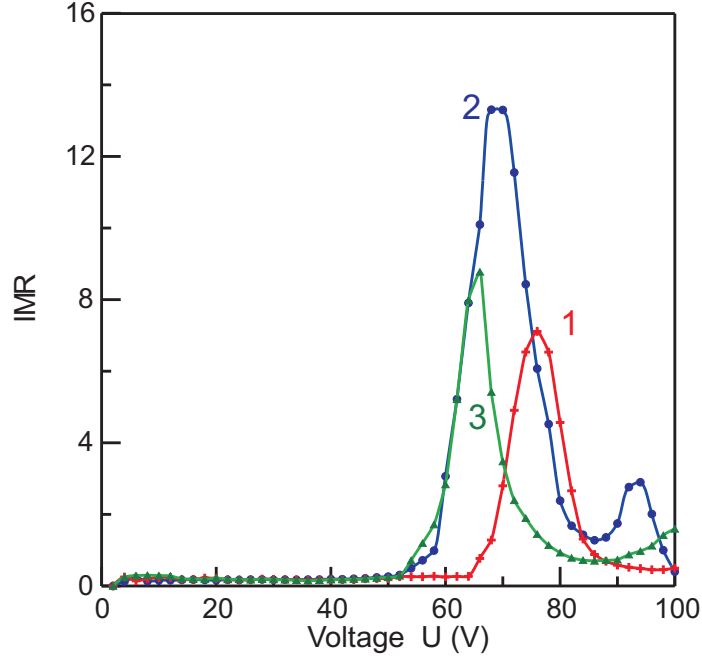}
\end{center}
\caption{Injection magnetoresistance ratio, IMR, of magnetic sensors
versus the applied voltage $U$ in the magnetic field $H$ = 2.1~kOe
for SiO${ }_2$(Co)/GaAs heterostructures with Co concentrations (1)
- 54 at.\%, (2) - 60 at.\%, (3) - 71 at.\%. } \label{Fig2}
\end{figure*}

The sizes of the samples in magnetic sensors were equal to $3\times
3\times 0.4$~mm. One contact was on the GaAs substrate, and the
other -- on the Au layer sputtered on the granular film. Magnetic
sensors are characterized by the injection magnetoresistance, which
is defined by the coefficient

\[{\mbox{\it IMR}}= \frac{R(H)-R(0)}{R(0)}=
\frac{j(0)-j(H)}{j(H)},\]

\noindent where $R(0)$ and $R(H)$ are the resistances of the SiO${
}_2$(Co)/GaAs heterostructure without a field and in the magnetic
field $H$, respectively; $j(0)$ and $j(H)$ are the current densities
flowing in the heterostructure in the absence of a magnetic field
and in the field $H$. The IMR ratio for SiO${ }_2$(Co)/GaAs
heterostructures with different Co concentrations versus the applied
voltage $U$ in the magnetic field $H$ = 2.1~kOe at room temperature
is shown in Fig. \ref{Fig2}. The magnetic field $H$ is parallel to
the film. According to \cite{Lut09}, for magnetic fields of high
values ($>$ 10~kOe) the IMR coefficient increases with the growth of
the applied voltage. In contrast to this, for low magnetic fields
the IMR reaches highest values in the region of the avalanche onset.
As one can see from Fig. \ref{Fig2}, in order to reach high
sensitivity of sensors it is need to apply the voltage in this
region.

\section{Field-effect transistor governed by magnetic field}

\begin{figure*}
\begin{center}
\includegraphics*[scale=.65]{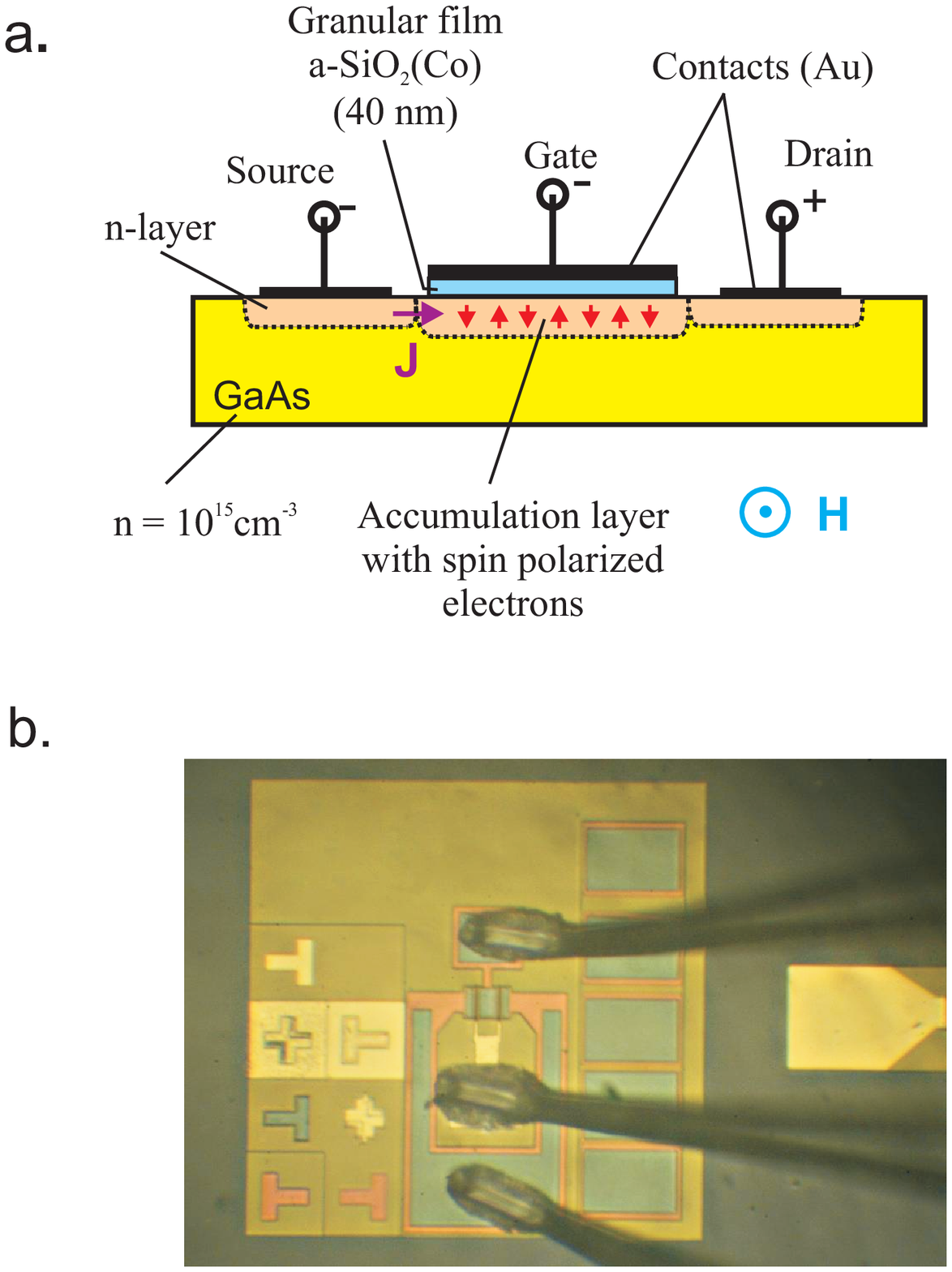}
\end{center}
\caption{Field-effect HEMT device with spin polarized electron
channel under gate electrode. (a) Schematic structure, (b) topology
of contacts. } \label{Fig3}
\end{figure*}

The field-effect HEMT device with a spin polarized electron channel
was developed on the base of the $n$-GaAs/AlGaAs heterostructures
(Fig. \ref{Fig3}). This device contains an amorphous granular SiO${
}_2$(Co) film with cobalt nanoparticles under the gate electrode.
The thickness of the film is equal to 40~nm. The SiO${ }_2$(Co) film
polarizes electron spins in the accumulation layer under the gate.

Current-voltage curves of field-effect devices have two different
parts~\cite{Pas}. If the voltage between the source and the drain
$U_{sd}$ is less than the saturation voltage $U_{sd}^{(sat)}$, then
the current-voltage curve is sub-linear and the current $J$ flowing
in the channel is written as

\begin{equation}
J=\frac{\mu Cb}{l}\left[(U_{gs}-U_{gs}^{(thr)})U_{sd}-\frac12
U_{sd}^2\right],\label{eq1}
\end{equation}

\noindent where $U_{gs}$ is the voltage between the gate and the
source, $U_{gs}^{(thr)}$ is the threshold voltage between the gate
and the source, when there is no current in the channel, $C$ is the
specific capacity between the gate and the channel, $\mu$ is the
electron mobility, $b$ and $l$ are the channel width and the length,
respectively. When $U_{sd}=U_{sd}^{(sat)}$, the channel becomes
blocked at the drain contact and an electrical field of high value
appears in this region. In this case, the saturation voltage is

\[ U_{sd}^{(sat)}=U_{gs}-U_{gs}^{(thr)}.\]

\noindent For $U_{sd}\geq U_{sd}^{(sat)}$ the drain current $J$ is
weakly dependent on the voltage $U_{sd}$ and the current-voltage
curve can be approximated by a line with a weak slope. In the first
approximation this sloping part of the current-voltage curve can be
written as

\begin{equation}
J=\frac{\mu Cb}{2l}(U_{gs}-U_{gs}^{(thr)})^2 ,\label{eq2}
\end{equation}

\begin{figure*}
\begin{center}
\includegraphics*[scale=.5]{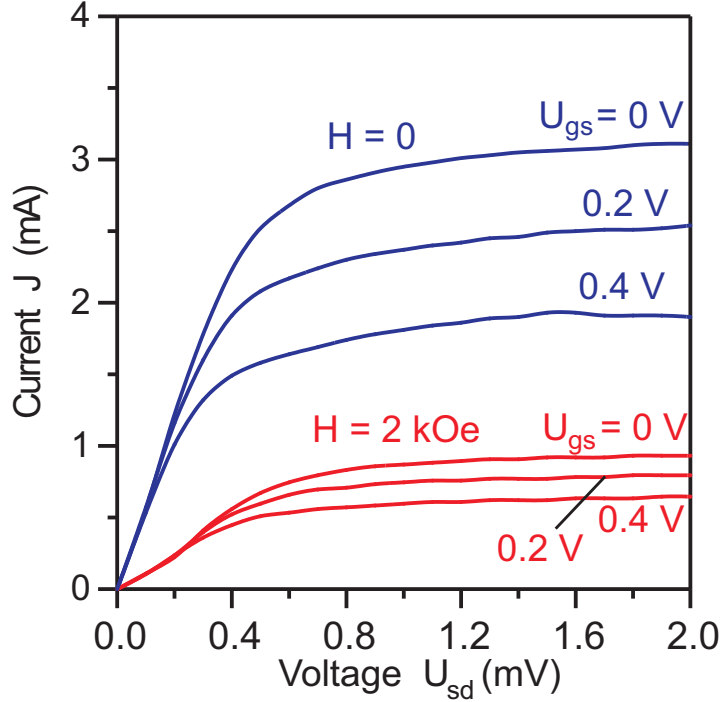}
\end{center}
\caption{Current-voltage dependences of the field-effect HEMT device
in the magnetic field $H$ = 2~kOe and without a magnetic field.
$U_{sd}$ is the drain voltage, $U_{gs}$  is the voltage at the gate
electrode. } \label{Fig4}
\end{figure*}

The current-voltage dependences of the developed field-effect HEMT
structure (Fig. \ref{Fig4}) contain two parts of the current-voltage
curve described by relations (\ref{eq1}) and (\ref{eq2}). The
saturation voltage $U_{sd}^{(sat)}$ is in the range 0.4 - 0.7~mV. As
one can see from Fig. \ref{Fig4}, the drain current $J$ of the
field-effect transistor presents strong dependence on the governed
external magnetic field $H$. The electron mobility $\mu$ in the
channel decreases with the growth of the applied magnetic field.
This property can be used in magneto-sensitive devices.

\section{Conclusions}

We study devices on the base of magnetic nanostructures containing
silicon dioxide films with cobalt nanoparticles SiO${ }_2$(Co) on
GaAs substrate -- magnetic sensors and field-effect transistor.
Magnetic sensors are based on the injection magnetoresistance effect
in the avalanche onset regime. The field-effect transistor contains
SiO${ }_2$(Co) film under gate. The both structures exhibit high
magnetic sensitivity at room temperature.

\section*{Acknowledgments}

The authors gratefully acknowledge the assistance of Dr. V.M.
Lebedev (PNPI, Gatchina, Leningrad region, Russia) for determination
of the film composition. This work was supported by the Russian
Foundation for Basic Research, grant 10-02-00516, and by the
Ministry of Education and Science of the Russian Federation, project
2011-1.3-513-067-006.

\end{document}